# DynaLog: An automated dynamic analysis framework for characterizing Android applications

Mohammed K. Alzaylaee[1], Suleiman Y. Yerima[1] and Sakir Sezer[1]
[1]Centre for Secure Information Technologies (CSIT)
Queen's University Belfast
Belfast, Northern Ireland
Email: {malzaylaee01, s.yerima, s.sezer}@qub.ac.uk

*Abstract*—Android is becoming ubiquitous and currently has the largest share of the mobile OS market with billions of application downloads from the official app market. It has also become the platform most targeted by mobile malware that are becoming more sophisticated to evade state-of-the-art detection approaches. Many Android malware families employ obfuscation techniques in order to avoid detection and this may defeat static analysis based approaches. Dynamic analysis on the other hand may be used to overcome this limitation. Hence in this paper we propose DynaLog, a dynamic analysis based framework for characterizing Android applications. The framework provides the capability to analyse the behaviour of applications based on an extensive number of dynamic features. It provides an automated platform for mass analysis and characterization of apps that is useful for quickly identifying and isolating malicious applications. The DynaLog framework leverages existing open source tools to extract and log high level behaviours, API calls, and critical events that can be used to explore the characteristics of an application, thus providing an extensible dynamic analysis platform for detecting Android malware. DynaLog is evaluated using real malware samples and clean applications demonstrating its capabilities for effective analysis and detection of malicious applications.

## I. INTRODUCTION

Google's Android operating system (OS) is increasingly widespread within the context of the modern market. This is due to the advent and rapid expansion of smartphones, with the number of smartphones shipped in the 2000s and early 2010s having tripled from 40 million to 120 million [1]. Android has an estimated market share of 70-80%, and it is the most popular operating system for smartphones and tablets. Since its release in 2008, over 50 billion total app downloads have been recorded [2]. In fact, it is expected that a shipment of one billion Android devices will be delivered in 2017 [3]. However, the popularity and growth of the Android OS has exposed it to the increasing threat of malware. In preparation for this significant shipment of Android devices, cyber criminals have expanded their activities. This has resulted in active research and development concerning Android malware in an effort to protect users.

The increased growth of the Android platform has highlighted the growing need for effective solutions to address the spread of mobile malware. The problem has worsened, due to the rapid evolution of mobile malware and its ability to avoid existing detection methods [4]. Furthermore, since the summer of 2010, there has been a 400% increase in Android-based malware. Moreover, the total number of Android malware samples exceeded 5 million in 2014 [5]. Therefore, there is a need to find new solutions to this growing problem, since traditional signature-based antivirus solutions are not effective especially against zero-day malware.

In recent years, several approaches have been investigated to improve the detection of Android malware. Many of the approaches have been based on static analysis while others employ dynamic analysis. Static analysis tools/frameworks such as Androguard [6], RiskRanker [7], APKinspector [8], Comdroid [9], etc. have been proposed, but these are vulnerable to the limitations of static analysis, i.e. detection avoidance by sophisticated obfuscation techniques, run-time loading of malicious payload etc. Several dynamic analysis based tools and platforms have also been proposed such as Taintdroid [10], Andromaly [11], Droidbox [12], DroidScope [13] etc. Some of these tools have been developed for a limited analysis scope. For example, Taintdroid, designed to detect only data leakage from an application. Other platforms, Droidbox for example, allow for dynamic analysis but only by manual means and hence cannot provide automated mass screening of apps without modification. Other dynamic analysis tools are either closed source or can only be accessed by submitting apps online for analysis, which can also limit automated mass analysis of apps by researchers/analysts.

Hence in order to overcome the aforementioned limitations we are motivated to design and develop a platform for automated dynamic analysis of Android applications named DynaLog. DynaLog is motivated by the need for the capability to automatically analyse massive amounts of apps and extract extensive characterization features that can enable us to gain insight into the dynamic behaviours of the apps. Furthermore, we want to be able to utilize these features to identify and isolate those applications that might exhibit malicious behaviour. In summary, the main contributions of this paper are as follows:

- We present DynaLog, a dynamic analysis framework to enable automated analysis of Android applications. DynaLog is built upon existing open source tools thus providing an extensible framework that enables a wide-ranging scope for dynamic analysis of Android apps.

- The framework components and features are described in detail. Furthermore, we discuss how existing open source tools such as Droidbox have been leveraged to build DynaLog and enable new app characterizing features that do not exist in most currently available dynamic analysis platforms.

- We present extensive experimental evaluation of DynaLog using real malware samples and clean applications in order to validate the framework and measure its capability to enable identification of malicious behaviour through the extracted behavioural features.

The remainder of the paper is organized as follows. Section II discusses the DynaLog framework. Section III presents evaluation experiments and discussions of the results. Section IV discusses related work while section V is the conclusion of the paper.

## II. DYNALOG FRAMEWORK

As mentioned earlier, DynaLog is motivated by the need for automated analysis of massive amounts of apps using dynamic analysis to help identify apps with malicious behaviour. Hence, DynaLog is designed automatically to accept a large number of apps, launch them serially in an emulator, log several dynamic behaviours and extract these for further processing. DynaLog has several components as shown in Figure 1. These include: a) Emulator-based analysis sandbox b) APK instrumentation module c) Behavior/features logging and extraction d) Application trigger/exerciser e) Log parsing and processing scripts.

### A. Emulator-based analysis sandbox

Most dynamic analyses require a sandboxed environment to run and analyse the applications under test. The DynaLog framework utilizes DroidBox [12] an open source tool that can be used to extract some high level behaviours and characteristics by running the app on an Android device emulator or Android Virtual Device (AVD). DroidBox extracts these behaviours from the logs dumped by *logcat* while the application is running on the AVD. It uses Androguard to extract static meta-data relating to the app and also utilizes Taintdroid for data leakage detection. Because DroidBox was the first open source dynamic analysis sandbox, it has been used as a building block for several dynamic analysis frameworks such as Sandroid [14], MobileSandbox [15] and Andrubis [16]. Table 1 shows the high level behaviours (features) that are available with DroidBox.

DynaLog was developed within the Santoku Linux environment because this Linux distribution is aimed at providing tools and utilities for Android security analysis. Santoku has the tools that DroidBox depends on to function properly including the Android platform tools such as *adb, logcat, Android emulator images* etc. Since DynaLog framework obtains the DroidBox-based default features shown in Table 1, from an emulator, it was necessary to enhance the emulator image as much as possible by changing some of its properties to bring it closer to a real device as possible (from the viewpoint of an application under analysis). This is because some malware are known to hide their malicious behaviour if through fingerprinting, they discover that they are installed and running in an emulator. The following have been applied to enhance the DynaLog framework sandbox emulator:

TABLE I. DEFAULT DROIDBOX FEATURES INCORPORATED INTO DYNALOG

| Feature Abbreviation | Feature Description |
|---|---|
| Hashes | Hashes for the analysed package |
| Opennet + Recvnet + Sendnet | Connections made with particular networks |
| Accessedfiles + Fdaccess | Reading and writing files |
| Servicestart | Started services |
| Dexclass | Loaded classes through DexClassLoader |
| Dataleaks | Information leaks via the network, file, and SMS |
| Enfperm | Circumvented permissions |
| Cryptousage | Cryptographic operations performed using the Android API |
| Recvaction | Listing broadcast receivers |
| Sendsms + Phonecalls | Sent SMS and phone calls |

- Because some applications can hide their malicious behaviour when running in an emulator, modifications were made to the IMEI, IMSI, Sim_Serial number, and phone number. For example, the default emulator IMEI number '000000000000000', was changed to a real IMEI number, '122XXX62XXX5532' using a hex editor to modify the property within the emulator image.

- Contact information was added to the emulator's Android.contact using the push command, '$ adb push contacts.vcf /sdcard/contacts.vcf'.

Even though DroidBox provides some features for characterizing applications, it does not provide the ability to log API calls. Android has hundreds of (Android and Java based) API calls that can be traced when the app is running on the emulator. Many of these API calls can be very helpful in identifying malicious behaviour and also in providing better characterization of applications in general. Also, the DroidBox features do not provide information that is granular enough or provides enough context for the behaviour. For example the *Recvaction* feature does not break down the broadcast events received. Also the Dexclass feature does not provide enough context of the loaded class. These limitations would make it more difficult to identify malware or malicious behaviour effectively.

Within the DynaLog framework new (granular) features have been enabled by extracting lower level features from the higher level *Recvaction* feature available in DroidBox. These features are represented by dozens of events/actions called by *Intents* within the application being analysed. In DynaLog we provide the capability to log and extract these events which can be used as features to characterize apps and potentially distinguish malware from benign applications.

### B. APK instrumentation module

As mentioned earlier, the open source DroidBox tool does not have the capability to extract and log API calls which are useful characteristics for dynamic analysis. Hence, we added an instrumentation module to DynaLog in order to enable API calls to be monitored, logged and extracted. The instrumentation module leverages APIMonitor, an open source tool that allows Android APKs to be instrumented using (smali and baksmali) assembler/disassembler tools. APIMonitor was developed by Yang at GSoC 2012 [12]. Leveraging APIMonitor allows DynaLog to be able to instrument an APK to monitor any API call available within Android and/or Java that developers can use to develop the apps. But since our main goal is to detect malicious behaviour we will be mostly interested in API calls that are commonly used by malware.

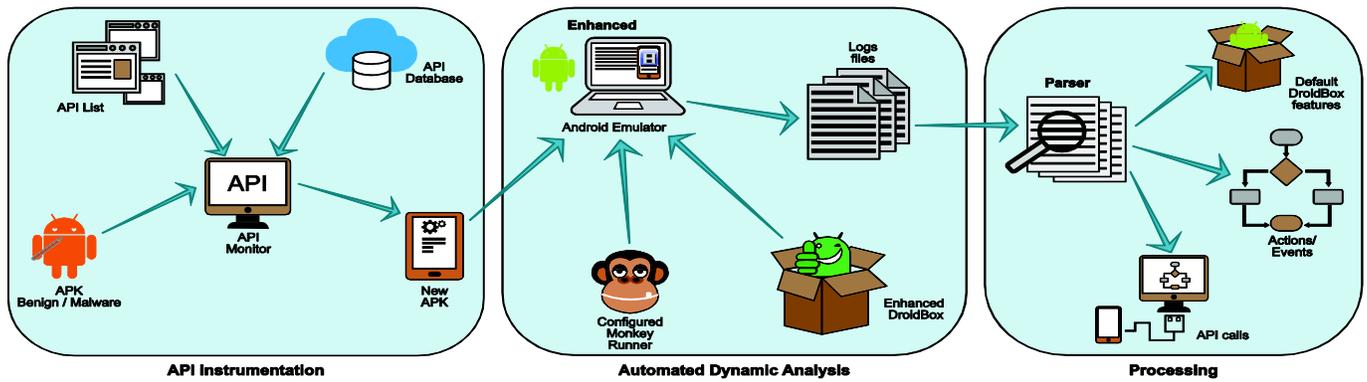

Fig. 1. DynaLog architecture

The instrumentation process involves reverse engineering the dex file using baksmali and inserting signatures that can be used to monitor the presence of an API call's class or methods within the log of the emulator while the app is running. As shown in Figure 1, the API calls instrumentation signatures are provided as a 'list' to the APIMonitor tool, and DynaLog automates the process of inserting the signatures to each application to be analysed in the first step within the overall analysis process shown in Figure 1. Examples of API call logs that DynaLog can extract after instrumenting an app are shown in Figure 2 below. The first one shows an API call that that sends via SMS the message '532711' to the destination number '1782'. The second example shows another API call to execute a process.

```
LAndroid / telephony /SmsManager;−> sendTextMessage ( Ljava / lang / String ;=
1782 | Ljava / lang /String;= null| Ljava / lang / String;= 532711 | LAndroid /app/
Pending Intent ;= null | LAndroid /app/ Pending Intent ;= null )V
```

```
Ljava / lang /Runtime;−>exec ( [ Ljava / lang / String ;={ / data / data / org . zen
though t . flashrec / cache / asroot , / data / data / org.zenthought.flashrec / cache /
explXXXXXX, / data / data / org.zenthought.flashrec / cache / dump image , recovery
, /mnt/ sdcard / recovery −backup . img }) Ljava / lang / Process ;= Process [ id=541] )
```

Fig. 2. API calls from an instrumented application

## C. Behaviour logging and extraction

DynaLog implements capability to extract specific log entries that correspond to monitored behaviours or API call signatures. If the extracted log entry is a high level property that could be further dissected into lower level features, the extracted entry is further parsed to capture the lower level features. Indeed, this is the case with the *Recvaction* log entry provided by DroidBox.

## D. Application trigger/exerciser

Android applications typically have a main application launcher activity which is usually the first screen that users interact with or see when an application is launched. As the user interacts with the screens, several paths are traversed to invoke its functionalities. With automated dynamic analysis, code coverage is an issue because the apps have to be artificially stimulated and it is often not possible to invoke all the application's paths during analysis. DroidBox includes MonkeyRunner by default. MonkeyRunner is an application exerciser that comes as one of the Android platform tools with the Android SDK and is provided for automated testing of applications. It sends random events such as 'touch screens', 'swipes', 'presses' etc. to an application under test. DroidBox only invokes the main application launcher activity by default. Thus, in order to improve code coverage for the DynaLog framework, we included capability to *invoke all the activities and services* that are present within an application. Presently DynaLog also uses MonkeyRunner to exercise the applications (with up to 3000 random events). In future we plan to incorporate a more advanced exerciser to ensure greater application path traversal.

## E. Log parsing and processing

In DynaLog, extracted features are properly formatted into readable output reports for each application being analysed with the automated framework. This allows the features present in each application to be overviewed at a glance or processed further by a classification engine.

### III. DYNALOG FRAMEWORK EVALUATION

In order to evaluate our proposed framework, we performed several experiments to investigate its capabilities using both benign applications and malware samples. We used 1226 real malware samples from 49 families of the Malgenome Project malware dataset. The families and their corresponding numbers are shown in Table II. Furthermore, a set of 1000 internally vetted benign APKs from McAfee Labs were utilized. Therefore, a total of 2226 malware and benign applications were used to conduct the experiments. Out of these, we were able to analyse 970 malware samples and 970 benign samples due to some of the applications not executing properly.

TABLE II. MALWARE FAMILIES USED AND THEIR NUMBERS

| Family | No. of samples | Family | No. of samples |
|---|---|---|---|
| DroidKungFu3 | 309 | GPSSMSSpy | 6 |
| AnserverBot | 187 | HippoSMS | 4 |
| BaseBridge | 122 | GingerMaster | 4 |
| DroidKungFu4 | 96 | DroidKungFuSapp | 3 |
| Geinimi | 69 | TapSnake | 2 |
| Pjapps | 58 | Crusewin | 2 |
| KMin | 52 | Nickyspy | 2 |
| GoldDream | 47 | RogueLemon | 2 |
| DroidDreamLight | 46 | SMSReplicator | 1 |
| DroidKungFu1 | 34 | Walkinwat | 1 |
| DroidKungFu2 | 30 | Endofday | 1 |
| ADRD | 22 | GGTracker | 1 |
| YZHC | 22 | GamblerSMS | 1 |
| DroidDream | 16 | Lovetrap | 1 |
| jSMSHider | 16 | Zitmo | 1 |
| Zsone | 12 | CoinPirate | 1 |
| zHash | 11 | DogWars | 1 |
| Plankton | 11 | NickyBot | 1 |
| SndApps | 10 | DroidCoupon | 1 |
| BgServ | 9 | DroidDeluxe | 1 |
| RogueSPPush | 9 | Spitmo | 1 |
| Gone60 | 9 | DroidKungFuUpdate | 1 |
| Asroot | 8 | FakeNetflix | 1 |
| BeanBot | 8 | Jifake | 1 |
| FakePlayer | 6 | | |

### A. Experiment 1: Evaluating high level behaviour features

As discussed in the previous section, DynaLog initially had high level behaviour features shown in Table I, which are obtained from the DroidBox component. In a preliminary experiment we tested these high level features using 106 APKs—53 benign and 53 malicious—for three minutes each by configuring DynaLog to enable only these default DroidBox based features. The results are shown in Figure 3, and they illustrate a lack of detailed information that can be used to distinguish between malware and benign applications effectively. Thus from this figure, it is clear that the high level behaviours captured by the DroidBox properties are neither extensive nor granular enough to provide good classification of applications.

### B. Experiment 2: Evaluating extended features and sandbox enhancements within DynaLog

As mentioned before, DynaLog enables new granular features by extracting further properties from the high level behaviour properties that DroidBox provides. In particular, Recvaction is used to extract several 'events' and 'actions' shown in Table III. Figure 4 shows sample output from Recvaction. From this, DynaLog can extract the BOOT_COMPLETED and UMS_DISCONNECTED events for the application. These are new features defined in DynaLog which other dynamic analysis framework do not have.

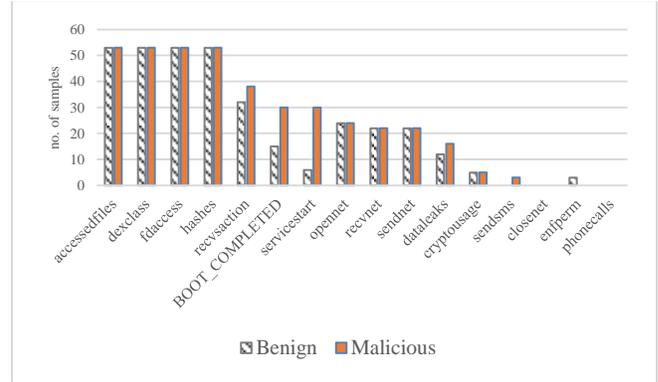

Fig 3. Behaviour logs from 106 APKs using DynaLog configured to enable only the default DroidBox features.

```
"recvsaction": {
"com.google.ssearch.Receiver": "Android.intent.action.BOOT_COMPLETED"
"com.Android.view.custom.BaseABroadcastReceiver":
"Android.intent.action.UMS_DISCONNECTED"
},
```

Fig. 4. Sample of the output from 'recvsaction'

The other set of extended features available from DynaLog are the API call traces. These are from the instrumentation module described in the previous section. An extensive API signature list has been included to track API calls that are likely to be found in malware applications. This signature list includes those for Telephony Manager API through which the following DynaLog features can be tracked: 'device_ID', 'subscriber_ID', 'lineNumber', 'SimSerialNumber', and 'SimOperatorNumber'. These features are crucial for detecting malicious applications.

Figure 5 illustrates the impact of granular properties (i.e. events) when both benign and malware samples were tested on the extended features of DynaLog. It can be seen that some of the features allow for better separation of malware from benign applications. 'PHONE_STATE' and 'BOOT_COMPLETED' were observed more frequently with the malware samples. 60% of the malware samples listened for the BOOT_COMPLETED event, whereas only 15% of the benign samples that were tested did. Even though 'PHONE_STATE' was used by almost half of the benign samples, over 90% of the malware samples logged this event.

TABLE III. ANDROID EVENTS AND ACTIONS RELATED TO MALWARE [4]

| Event | Abbreviation | Event | Abbreviation | Event | Abbreviation |
|---|---|---|---|---|---|
| BOOT_COMPLETED | BOOT (Boot Completed) | SMS_RECEIVED WAP_PUSH_RECEIVED | SMS (SMS/MMS) | ACTION_MAIN | MAIN (Main Activity) |
| PHONE_STATE NEW_OUTGOING_CALL | CALL (Phone Events) | UMS_CONNECTED UMS_DISCONNECTED | USB (USB Storage) | CONNECTIVITY_CHANGE PICK_WIFI_WORK | NET (Network) |
| PACKAGE_ADDED PACKAGE_REMOVED PACKAGE_CHANGED PACKAGE_REPLACED PACKAGE_RESTARTED PACKAGE_INSTALL | PKG (Package) | ACTION_POWER_CONNECTED ACTION_POWER_DISCONNECTED BATTERY_LOW BATTERY_OKAY BATTERY_CHANGED_ACTION | BATT (Power/Battery) | USER_PRESENT INPUT_METHOD_CHANGED SIG_STR SIM_FULL | SYS (System Events) |

In order to investigate the impact of the sandbox enhancements implemented to make the emulator seem more realistic to applications, we employed 31 samples from the DroidKungFu1 family. The results are shown in Table IV. We observed that for some features such as DeviceId, SubscriberId, RuntimeExec, SimSerialNumber, getLineNumber there was a marked difference in the results observed before enhancement and after enhancement.

TABLE IV. SUBSET OF RESULTS FROM DROIDKUNGFU1 SAMPLES

| Properties | Result before sandbox enhancement | Result after sandbox enhancement |
|---|---|---|
| getDeviceId (TelephonyManager) | 10 | 14 |
| getSubsriberId (TelephonyManager) | 3 | 9 |
| getSimSerialNumber (TelephonyManager) | 3 | 9 |
| getLine1Number (TelephonyManager) | 1 | 8 |
| Runtime.exec() (Excuting process) | 1 | 10 |

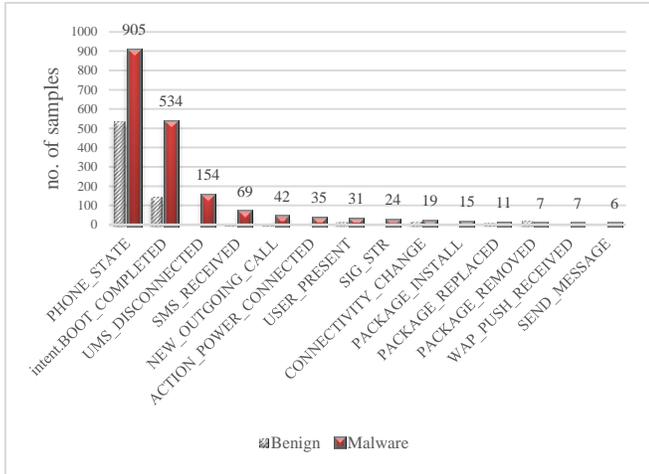

Fig. 5. Events observed from the extended feature set of DynaLog

These results of table IV confirm observations made in previous work such as [17]-[21] about the limitations of dynamic analysis and illustrates the necessity to incorporate more realistic properties into emulator environments when attempting to detect malware.

### C. Results of malware and benign samples comparison

In this section we present the results of comparing benign applications to malware applications using the DynaLog framework. In this experiment the available features from the feature sets: high level behaviour, granular events, API calls, were all enabled. Table V shows the results obtained for 44 features from DynaLog.

Table V shows that 905 APKs in malware samples logged the 'PHONE_STATE' event. This was the feature most observed within the malware samples. 'servicestart' ranked second, and was called by 840 APKs. The collection of phone information in order to send to a remote server is a considerable concern, and malware often seeks to do so, as shown by 'getDeviceId', 'getSubscriberId', 'getSimSerialNumber', NetworkOperator, 'Line1Number', and 'getSimOperator'. The results in Table V illustrates the capability of the features logged by DynaLog to characterize applications and potentially separate malicious applications from benign ones.

TABLE V. COMPARISON RESULTS OF 970 BENIGN AND 970 MALWARE SAMPLES

| | Top Extracted Features | Benign | Malware | % Benign | % Malware |
|---|---|---|---|---|---|
| 1 | PHONE_STATE | 537 | 905 | 55.36 | 93.29 |
| 2 | servicestart | 603 | 840 | 62.16 | 86.59 |
| 3 | PackageManager | 441 | 601 | 45.46 | 61.95 |
| 4 | intent.BOOT_COMPLETED | 150 | 534 | 15.46 | 55.05 |
| 5 | Process | 287 | 480 | 29.58 | 49.48 |
| 6 | opennet | 295 | 471 | 30.41 | 48.55 |
| 7 | checkPermission | 169 | 456 | 17.42 | 47.01 |
| 8 | sendnet | 250 | 421 | 25.77 | 43.40 |
| 9 | recvnet | 244 | 418 | 25.15 | 43.09 |
| 10 | getInstance | 279 | 417 | 28.76 | 42.98 |
| 11 | deviceId | 229 | 367 | 23.60 | 37.83 |
| 12 | getMethod | 256 | 358 | 26.39 | 36.90 |
| 13 | parse | 190 | 316 | 19.58 | 32.57 |
| 14 | digest | 221 | 288 | 22.78 | 29.69 |
| 15 | dataleaks | 147 | 282 | 15.15 | 29.07 |
| 16 | getClass | 120 | 226 | 12.37 | 23.29 |
| 17 | SubscriberId | 40 | 225 | 4.123 | 23.19 |
| 18 | cryptousage | 93 | 219 | 9.58 | 22.57 |
| 19 | SimSerialNumber | 13 | 212 | 1.34 | 21.85 |
| 20 | lineNumber | 33 | 190 | 3.40 | 19.58 |
| 21 | start | 176 | 176 | 18.14 | 18.14 |
| 22 | NetworkOperator | 44 | 171 | 4.53 | 17.62 |
| 23 | UMSDISCONNECTED | 0 | 154 | 0 | 15.87 |
| 24 | ContentResolver | 55 | 153 | 5.67 | 15.77 |
| 25 | connect | 50 | 105 | 5.15 | 10.82 |
| 26 | getApplicationInfo | 48 | 91 | 4.94 | 9.38 |
| 27 | SimOperator | 26 | 85 | 2.68 | 8.76 |
| 28 | runtime.exec | 40 | 70 | 4.12 | 7.21 |
| 29 | initCipher | 67 | 70 | 6.90 | 7.21 |
| 30 | getInstance | 57 | 70 | 5.87 | 7.21 |
| 31 | SMSRECEIVED | 7 | 69 | 0.72 | 7.11 |
| 32 | SecretKey | 46 | 64 | 4.74 | 6.59 |
| 33 | SimCountryIso | 22 | 44 | 2.26 | 4.53 |
| 34 | NEW_OUTGOING_CALL | 11 | 42 | 1.13 | 4.32 |
| 35 | ACTION_POWER_CONNECTED | 0 | 35 | 0 | 3.60 |
| 36 | USER_PRESENT | 22 | 31 | 2.26 | 3.195 |
| 37 | SIG_STR | 0 | 24 | 0 | 2.47 |
| 38 | sendsms | 0 | 18 | 0 | 1.85 |
| 39 | getLastKnownLocation | 12 | 17 | 1.23 | 1.75 |
| 40 | openOrCreateDatabase | 15 | 16 | 1.54 | 1.64 |
| 41 | PACKAGE_INSTALL | 1 | 15 | 0.103 | 1.546 |
| 42 | WAP_PUSH_RECEIVED | 3 | 7 | 0.309 | 0.721 |
| 43 | phonecalls | 0 | 6 | 0 | 0.618 |
| 44 | SEND_MESSAGE | 0 | 6 | 0 | 0.618 |

## D. Limitations of the DynaLog framework

DynaLog suffers from the same limitations that are well documented for dynamic analysis [17] [22] [23]. Hence we intend to continue to improve the framework to overcome these limitations. For apps that fail to log any output for dynamic analysis we may resort to static analysis based features in future. Presently, DynaLog does not log output from native code. Nevertheless, DynaLog incorporates API calls such as *system.LoadLibrary* that can indicate when calls are being made to native code libraries. DynaLog is designed to enable extension of features, as such system call related features could also be incorporated in future.

## IV. RELATED WORKS

There are two fundamental practices in malware analysis: static analysis, which involves examining malware without running it; and dynamic analysis, which involves examining the malware whilst it runs. Several tools and frameworks have been proposed for detecting Android malware, and most of these can be categorized in this way.

### A. Android Static Analysis

Several static analysis approaches and techniques have been proposed in the literature. Androguard [6] is a static analysis tool that can disassemble and decompile Android applications. ComDroid [9] is another static analysis tool for Android applications that detects vulnerabilities in communication. APKInspector [8] is a powerful GUI tool based on static analysis that can display a control-flow graph, application meta-data, Java source code, Dalvik bytecode, etc. Yerima et al. [24] presented an approach to detecting Android malware based on Bayesian classification models, using 58 features—including API calls, system commands, and intents—to classify Android applications as 'malicious' or 'benign'. Further study was done on the framework by adding features driven from permissions in order to study the impact of Bayesian-based classification [25]. Yerima et al. [26] also proposed parallel classifiers for the features extracted from Android applications, and compared multiple parallel classification combinations in order to improve the detection rate.

Although static analysis frameworks provide valuable insight into Android application behaviour, there are several limitations to static analysis, including a vulnerability to sophisticated detection-avoidance techniques and update attacks. Some malware apply sophisticated techniques, such as obfuscation and polymorphic techniques that make it difficult to detect [27].

### B. Android Dynamic Analysis

#### a) Machine-learning frameworks

Shabtai et al. [11] provide a dynamic analysis framework called Andromaly, which applies several different machine-learning algorithms to classify Android applications. However, they evaluated their techniques with four self-written malware applications, and it is unclear how they measured the detection performance.

MADAM, a multi-level anomaly detector for Android malware [28], is similar to Andromaly. MADAM monitors 13 features at the kernel level and the user level. A machine-learning algorithm was used to classify the applications. However, they tested the framework on only 2 malware samples with 50 benign samples. Therefore, their results are limited to a small dataset. Crowdroid [29] is a machine-learning framework based on a system called Strace. Crowdroid tests in the cloud, and in this way it differs from MADAM and Andromaly, both of which perform testing within the device. Furthermore, the evaluation was done using only two self-written malware samples.

#### b) Open-source frameworks

TaintDroid is a revised version of the Android OS (version 2.1) that was introduced by Enck et al. in October, 2010 [10]. This platform is able to monitor tainted data during runtime in order to identify private-information leaks. TaintDroid was implemented based on the Dalvik virtual machine. A number of studies [11], [30], [31], [32] have used TaintDroid for Android malware detection. The drawback to this platform is that it can only detect data leakage, whereas other malicious runtime behaviour can evade it, as demonstrated by Scrubdroid [33]. TaintDroid has been introduced into other dynamic analysis platforms, such as DroidBox and AppsPlayground [34], for data-leakage detection.

Lantz developed DroidBox at the Google Summer of Code (GSoC) in 2011 [12]. The platform employs an integrated system, containing TaintDroid with a modification of Android's core libraries. Moreover, DroidBox provides a visual illustration of the analysis results, and it installs automatically and runs as an Android Virtual Device (AVD). DroidBox is the first open-source dynamic analysis platform for Android. Therefore, it has been employed as a base system by various dynamic analysis platforms, such as Andrubis [16], Mobile-Sandbox [15], and SandDroid [14]. Furthermore, DroidBox only considers a limited set of behaviour, and it is restricted to one kernel version. DroidScope [13] is a wide-ranging dynamic binary-control mechanism for Android based on virtual machine (VM) analysis. It was introduced in August, 2012, as a modification of Dalvik's 'traces', and it is attached to an emulator.

#### c) Online services

Google introduced Bouncer in February, 2012 [35]. Bouncer is a dynamic analysis platform that is used to scan submitted apps for potentially malicious behaviour. During Summercon in June, 2012, Oberheide and Miller presented their research regarding Bouncer [36]. They concluded that Bouncer could easily be evaded by malware applications.

Andrubis is another dynamic analysis platform for Android applications, introduced in June, 2012, by the International Secure Systems Lab [16]. This framework was the first publicly online platform for dynamic analysis of Android applications. However, the code is not publicly available. Moreover, Andrubis cannot be used for large-scale analysis. Therefore, only a few applications can be uploaded at a time.

CopperDroid was presented by Reina et al. in April, 2013 [17]. The operating system for this platform is similar to that of

DroidScope—i.e. both use VMI to track system call information regarding analysed applications. The application also allows its users to use an online portal to submit applications for analysis. Tracedroid is another free online analysis service that analyses applications using dynamic and static analysis [37].

Like some of the existing dynamic analysis frameworks such as MobileSandbox, DynaLog uses the open source DroidBox as one of its building blocks. However, it introduces new granular features (i.e. events/actions). DynaLog is also an extensible framework that enables automated mass dynamic analysis of Android applications.

## V. CONCLUSIONS

In this paper, we presented DynaLog a framework that enables automated mass dynamic analysis of applications in order to characterize them for analysis and potential detection of malicious behaviour. DynaLog was built using existing open source tools and motivated by the need for an automated analysis platform since most existing frameworks are either closed source or allow only intermittent submissions of application online for analysis. DynaLog incorporates an emulator-based analysis sandbox based on DroidBox and implements an instrumentation module that allows API calls signatures to be embedded in applications so as to log various potentially malicious behaviour enabled through some API classes and methods. We have performed several experiments to evaluate the framework and the results presented in this paper demonstrates its capabilities and effectiveness as a platform that can be used for mass detection of sophisticated Android malware. For future work we intend to develop and couple classification engines that can utilize the extensive features of DynaLog for accurate identification of malware samples. Furthermore, we intend to enhance the framework to improve its robustness against anti-analysis techniques employed by some malware whilst also incorporating new feature sets to improve the overall analysis and detection capabilities.

## ACKNOWLEDGMENT

We gratefully acknowledge the sponsorship of the PhD project upon which some of this work is based, by Umm Al-Qura University, Saudi Arabia, through the Saudi Arabian Cultural Bureau in London, UK.